\documentclass[aps,pra,twocolumn,superscriptaddress]{revtex4-1}
\pdfoutput=1
\usepackage[utf8]{inputenc} 
\usepackage[english]{babel} 
\usepackage[T1]{fontenc}
\usepackage{hyperref}
\usepackage{graphicx}  
\usepackage{natbib}
\setcitestyle{numbers,sort,compress}
\usepackage{subfigure}
\usepackage{amssymb}   
\usepackage{amsmath}
\usepackage{mathtools}
\usepackage{color}
\usepackage{tabulary}
\newcolumntype{K}[1]{>{\centering\arraybackslash}p{#1}}
\usepackage[normalem]{ulem}

\newcommand{\bra}[1]{\langle #1|}
\newcommand{\ket}[1]{| #1 \rangle}

\begin{document}

\title{Variational quantum simulation of long-range interacting systems}

\author{Chufan Lyu}
\affiliation{Institute of Fundamental and Frontier Sciences, University of Electronic Science and Technology of China, Chengdu 610051, People's Republic of China}

\author{Xiaoyu Tang}
\affiliation{Institute of Fundamental and Frontier Sciences, University of Electronic Science and Technology of China, Chengdu 610051, People's Republic of China}

\author{Junning Li }
\affiliation{Central Research Institute, Huawei Technologies, Shenzhen 518129, People's Republic of China}

\author{Xusheng Xu}
\affiliation{Central Research Institute, Huawei Technologies, Shenzhen 518129, People's Republic of China}

\author{Man-Hong Yung}
\affiliation{Central Research Institute, Huawei Technologies, Shenzhen 518129, People's Republic of China}
\affiliation{Department of Physics, Southern University of Science and Technology, Shenzhen 518055, People's Republic of China}
\affiliation{Shenzhen Institute for Quantum Science and Engineering, Southern University of Science and Technology, Shenzhen 518055, People's Republic of China}
\affiliation{Guangdong Provincial Key Laboratory of Quantum Science and Engineering, Southern University of Science and Technology, Shenzhen 518055, People's Republic of China}
\affiliation{Shenzhen Key Laboratory of Quantum Science and Engineering, Southern University of Science and Technology, Shenzhen 518055, People's Republic of China}

\author{Abolfazl Bayat}
\email{abolfazl.bayat@uestc.edu.cn}
\affiliation{Institute of Fundamental and Frontier Sciences, University of Electronic Science and Technology of China, Chengdu 610051, People's Republic of China}

\date{\today}

\begin{abstract}
Current quantum simulators suffer from multiple limitations such as short coherence time, noisy operations, faulty readout and restricted qubit connectivity in some platforms. Variational quantum algorithms are the most promising approach in near-term quantum simulation to  achieve practical quantum advantage over classical computers.  Here, we explore variational quantum algorithms, with different levels of qubit connectivity, for digital simulation of the ground state of long-range interacting systems as well as generation of spin squeezed states. We find that as the interaction becomes more long-ranged, the variational algorithms become less efficient, achieving lower fidelity and demanding more optimization iterations. In particular, when the system is near its criticality the efficiency is even lower. Increasing the connectivity between distant qubits improves the results, even with less quantum and classical resources. Our results show that by mixing circuit layers with different levels of connectivity one can sensibly improve the performance. Interestingly, the order of layers becomes very important and grouping the layers with long-distance connectivity at the beginning of the circuit outperforms other permutations. The same design of circuits can also be used to variationally produce spin squeezed states, as a resource for quantum metrology. 
\end{abstract}

\maketitle

\section{Introduction}\label{section:Introduction}

Simulating quantum many-body systems on classical computers rapidly becomes intractable due to exponential growth of the Hilbert space. In fact, a true simulation of quantum systems is only feasible on another quantum system, called quantum simulator, which is normally less complex and more controllable than the system of interest~\cite{georgescu2014quantum}. Thanks to recent advancements in quantum technologies, quantum simulators are now emerging in various physical systems, including cold atoms~\cite{bloch2012quantum,gross2017quantum,simon2011quantum,yang2020cooling}, superconducting devices~\cite{arute2020hartree,kandala2017hardware,chen2020demonstration,harrigan2021quantum,gong2021quantum,yan2019strongly,gong2021experimental,mi2021information}, ion-tarps~\cite{kokail2019self,hempel2018quantum,pagano2020quantum}, Rydberg atoms~\cite{saffman2010quantum,weimer2010rydberg,ebadi2021quantum,bernien2017probing} and optical systems~\cite{spring2013boson,tang2022generating,peruzzo2014variational}. However, current Noisy Intermediate-Scale Quantum (NISQ) simulators are far from being perfect~\cite{preskill2018quantum}. Limited qubit connectivity in some platforms, relatively short coherence time, imperfect initialization, noisy operation and faulty readout are typical problems in current devices. Each NISQ simulator suffers from one or several of these issues. 
Quantum advantage has already been achieved for sampling problems on different NISQ platforms~\cite{Arute2019,Zhongadvantage2020,ZhongGaussian2021,WuAdvantage2021,Madsen2022}. However, sampling has little practical application and thus a hotly debated subject is whether NISQ simulators can provide any practical advantage over classical computers~\cite{altman2021quantum}.
Variational Quantum Algorithms (VQA)~\cite{cerezo2021variational,bharti2021noisy} are among the most promising approaches to achieve such practical quantum advantage through dividing the complexity between a quantum simulator and a classical optimizer. So far, VQAs have been developed in solving various problems in quantum machine learning~\cite{biamonte2017quantum,arunachalam2017survey,ciliberto2018quantum,dunjko2018machine,farhi2018classification,schuld2019quantum}, optimization~\cite{farhi2014quantum,bravyi2020obstacles,liu2022layer}, many-body systems~\cite{cirstoiu2020variational,gibbs2021longtime,yuan2019theory,mcardle2019variational,heya2019subspace,BravoPrieto2020scalingof,Lyu2020accelerated,uvarov2020variational,okada2022identification}, metrology~\cite{meyer2021variational,meyer2021fisher,beckey2020variational,kaubruegger2019variational,koczor2020variational,ma2021adaptive,haug2021natural} and chemistry~\cite{cao2021larger,arute2020hartree,peruzzo2014variational,kandala2017hardware,nam2020ground}.

Variational Quantum Eigensolver (VQE) is one of the most established VQAs for generating the low-energy eigenstates, specially the ground state, of a many-body Hamiltonian~\cite{peruzzo2014variational,mcclean2016theory,strutt1894theory,Ritz1909}. In this algorithm, a parameterized quantum circuit is used to generate a complex quantum state from a simple input wave function. Then the average energy of the desired Hamiltonian is measured at the output of the parameterized quantum circuit. This measured average energy is fed into a classical optimizer to be iteratively minimized. At each iteration the classical optimizer provides a new update for the parameters of the quantum circuit and a new set of measurements on the quantum simulator provides a new estimation for the average energy. As the average energy reaches its global minimum the output of the quantum circuit simulates the ground state of the system. VQE has also been generalized for simulating excited states through addition of penalizing terms to the cost function~\cite{higgott2019variational,kuroiwa2021penalty,mcclean2016theory,ryabinkin2018constrained}, subspace-search VQE~\cite{nakanishi2019subspace} and exploiting symmetries~\cite{barkoutsos2018quantum,wang2009efficient,Lyu2020accelerated,seki2020symmetry,gard2020efficient,barron2021preserving,zhang2021shallow,lyu2022symmetry,Meyerexploiting2023}. The complexity of any VQA, including the VQE, can be quantified through both quantum and classical resources which are required to successfully accomplish the target task. Quantum resources can be evaluated through the minimum depth of the quantum circuit. Since the most challenging ingredient of any quantum circuit is the two-qubit entangling gate, e.g. controlled-not gate, one can naturally use the number of such gates as a measure of quantum resources. The classical resources are determined by the complexity of the optimization. This can be quantified by the number of tunable parameters in the circuit and the number of iterations which is required for convergence of the algorithm.

Long-range interactions are very common and most fundamental forces in nature are naturally long-ranged. In these systems the strength of interaction between two particles decays as $1/d^\alpha$, where $d$ is the distance and $\alpha$ is an exponent which controls the strength of interaction. Smaller values of $\alpha$ make the interaction more long-ranged. Coulomb (with $\alpha=1$), dipole-dipole (with $\alpha=3$) and van der Waals (with $\alpha=6$) interactions are just a few examples of long-range forces in nature. The presence of long-range interaction in a many-body system can lead to the emergence of several interesting phenomena~\cite{ruelle1968statistical,dyson1969existence,cardy1981one,lahaye2009physics,frerot2018multispeed,frerot2017entanglement,meinert2014observation,buyskikh2016entanglement} with rich phase diagrams~\cite{koffel2012entanglement}. In particular, the generation of spin squeezing~\cite{kitagawa1993squeezed,ma2011quantum} through long-range interactions has both fundamental and practical implications. Fundamentally, they can be used for testing the foundations of quantum mechanics~\cite{schmied2016bell,engelsen2017bell}. From a practical perspective, they can be used as resource for metrology purposes to achieve quantum enhanced precision~\cite{pezze2018quantum}.

An interesting subject to explore is the possibility of VQE simulation for generating the ground state of a long-range interacting system~\cite{kokail2019self,pagano2020quantum,you2021exploring} and spin squeezed states~\cite{kaubruegger2019variational,marciniak2021optimal}.  Thanks to the presence of long-range couplings, the ground state of such systems shows significant correlation between distant particles and more multipartite entanglement. So far, analog VQE simulations have been used for generating the ground state of long-range interacting systems~\cite{kokail2019self,pagano2020quantum}. However, in such simulations the quantum simulator should already support the same type of long-range interaction between its qubits which may not be feasible. It is thus highly desirable to implement such simulation on digital quantum simulators, as universal computing machines rather than non-universal analog ones.
Digital quantum simulators, in NISQ era, come with different levels of qubit connectivity and thus limited interactions between their qubits. In some physical setups, such as superconducting systems~\cite{arute2020hartree,kandala2017hardware,chen2020demonstration,harrigan2021quantum,gong2021quantum,yan2019strongly,gong2021experimental,mi2021information} or cold atoms in optical lattices~\cite{spring2013boson,tang2022generating,peruzzo2014variational}, the connectivity is determined by the geometry of the simulator and is normally restricted to nearest neighbors. In other platforms, such as ion-traps~\cite{kokail2019self,hempel2018quantum,pagano2020quantum} and Rydberg atoms~\cite{saffman2010quantum,weimer2010rydberg,ebadi2021quantum,bernien2017probing}, the connectivity is more versatile and in principle interactions can be induced between any pair of qubits. A natural open problem is whether the flexibility in qubit connectivity can help digital VQE simulation of the long-range interacting systems. Although, one may expect that longer-distance qubit connectivity should enhance the performance of the VQE simulation of long-range interacting systems, no demonstration still exist to quantitatively show this effect.

In this paper, we address this issue through exploring the VQE simulation of long-range XY model in transverse field. In this system, one can control the anisotropy $\gamma$ to change the Hamiltonian from Ising in transverse field to anisotropic XX model. Moreover, by tuning the parameter $\alpha$ one can control the strength of long-range interactions. In addition, by controlling the ratio between the exchange interaction and the transverse field one can vary the phase of the system. Therefore, these three control parameters produce a large class of models with a rich phase diagram to be explored. We perform digital VQE simulation for the ground state of the system across its phase diagram assuming different levels of qubit connectivity. We find that the VQE simulation gets harder as the interaction is more long-ranged, i.e. smaller $\alpha$. In addition, around the quantum phase transition achieving higher fidelity is more challenging and requires more iterations for the convergence of the classical optimizer. Our results show that having connectivity  between more distant qubits not only improves the fidelity but also  speeds up the convergence of the classical optimizer. Interestingly, grouping the circuit layers with long-distance connectivity at the beginning of the circuit can improve the VQE simulation in terms of both achievable fidelity and required optimization iterations.

The structure of the paper is as follows. We start by giving a brief review on VQE in Section~\ref{section:VQE}, showing the important ingredients of such algorithm. Then, in Section~\ref{section:model}, we introduce the long-range XY Hamiltonian which contains three individual parameters with which one can tune the long-range interaction, the ratio between the exchange interaction and the strength of the transverse field, as well as the anisotropy to change the model from Ising in transverse field to anisotropic XX model. In Section~\ref{section:simulation}, we provide VQE simulation for the ground state of the long-range Ising model in transverse field using different levels of qubit connectivity. We further explore the impact of qubit connectivity in Section~\ref{section:scaling}. In Section~\ref{section:generality}, we provide VQE simulation results for the ground state of various models to demonstrate the generality of our strategy for circuit design.
We then show that how our quantum circuits can be used for generating spin squeezing in Section~\ref{section:squeezing}. Finally, we conclude our results in Section~\ref{section:conclusion}.

\begin{figure}[t]
	\centering
	\includegraphics[width=\columnwidth]{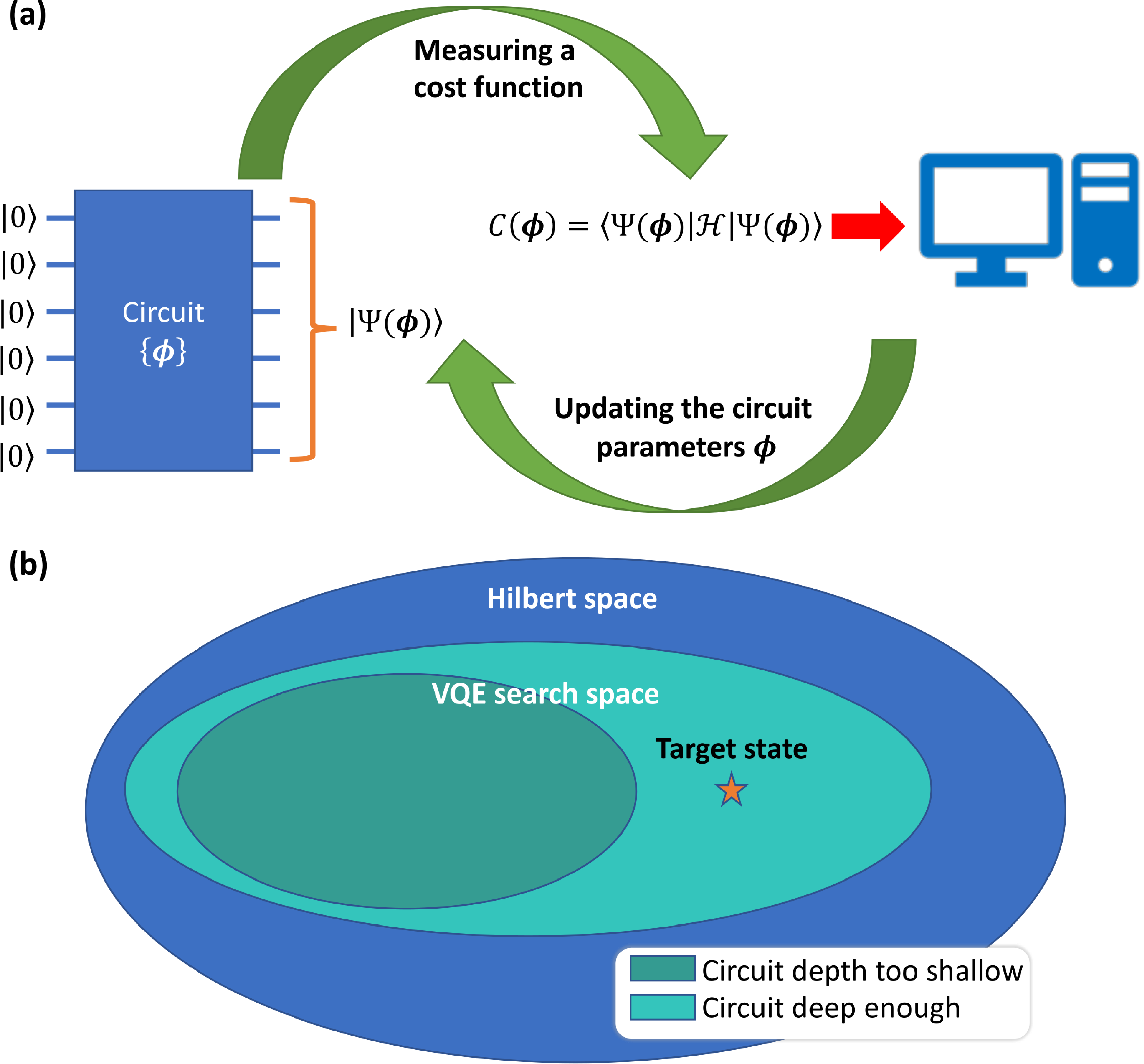}
	
	\caption{(a) The procedure of the VQE algorithm. In this algorithm, the output of the parameterized quantum circuit will be $\ket{\Psi(\boldsymbol{\phi})}$. A cost function $C(\boldsymbol{\phi})$ is then measured directly as the average of an observable $\mathcal{H}$ at the output of the quantum circuit. The circuit parameters $\boldsymbol{\phi}$ are updated iteratively via a classical optimizer to minimize the measured cost function $C(\boldsymbol{\phi})$. The procedure stops once an optimal set of circuit parameters $\boldsymbol{\phi}{=}\boldsymbol{\phi}^*$ is reached. (b) The schematic picture of the impact of circuit depth on the VQE algorithm. For a very shallow circuit, the span set $\ket{\Psi(\boldsymbol{\phi})}$ will not be able to include the target state. Therefore, the final output of the VQE algorithm $\ket{\Psi(\boldsymbol{\phi}^*)}$ cannot have a good approximation over the target state. As the depth of the quantum circuit increases, the span set $\ket{\Psi(\boldsymbol{\phi})}$ will eventually contain the target state, so that the target state can be well approximated if the global minimum is reached during the optimization. }
	\label{fig:vqe_schematic}
\end{figure}

\section{Variational Quantum Eigensolver}\label{section:VQE}

Here, we briefly review the VQE algorithm~\cite{peruzzo2014variational} for generating the ground state of a many-body system on a quantum simulator. 
Those who are familiar with the VQE algorithm can skip this section.
A parameterized quantum circuit implements an operation described by a unitary operator $\mathcal{U}(\boldsymbol{\phi})$, where $\boldsymbol{\phi}=(\phi_1,\phi_2,\cdots,\phi_L)$ represents the tunable parameters of local rotations in the quantum circuit. For a given initial state $\ket{\Psi_0}$, the output of the quantum circuit is given by $\ket{\Psi(\boldsymbol{\phi})}=\mathcal{U}(\boldsymbol{\phi})\ket{\Psi_0}$. By varying $\boldsymbol{\phi}$, the quantum state $\ket{\Psi(\boldsymbol{\phi})}$ can span some part of the Hilbert space. As the quantum circuit gets deeper, and thus the number of parameters $L$ increases, the spanned part enlarges until eventually it covers the whole Hilbert space. 
In the NISQ era, the goal is to keep the circuit as shallow as possible and yet being able to approximate the target state for a set of parameters $\boldsymbol{\phi}^*$. 
Designing such an optimal quantum circuit is a non-trivial task and highly depends on the underlying problem. For instance, for simulating the ground state of a many-body system, a circuit based on adiabatic evolution becomes extremely deep, well beyond the capability of currently available NISQ simulators~\cite{Lyu2020accelerated}. Variational quantum algorithms have been developed to achieve the required unitary operation on a shallow circuit. This is accomplished through dividing the complexity between a parameterized quantum circuit and a classical optimizer. 
As schematically shown in Fig.~\ref{fig:vqe_schematic}(a), the output of the quantum circuit will be $\ket{\Psi(\boldsymbol{\phi})}$.
The target state has to be formulated variationally as a minimum of a cost function $C(\boldsymbol{\phi})$ which is directly measured as the average of an observable $\mathcal{H}$ at the output of the quantum circuit such that $C(\boldsymbol{\phi})=\bra{\Psi(\boldsymbol{\phi}) } \mathcal{H} \ket{\Psi(\boldsymbol{\phi})}$. In the VQE algorithm, the observable $\mathcal{H}$ should be the Hamiltonian of the system and the cost function becomes the average energy~\cite{peruzzo2014variational}.
By classically optimizing the cost function $C(\boldsymbol{\phi})$ with respect to parameters $\boldsymbol{\phi}$, one can iteratively reach to an optimal set of parameters $\boldsymbol{\phi}=\boldsymbol{\phi}^*$. Provided that the span set $\ket{\Psi(\boldsymbol{\phi})}$ includes the target state, i.e. the ground state $\ket{GS}$, then the output of the quantum circuit $\ket{\Psi(\boldsymbol{\phi^*})}$ well approximates the target state, namely $\ket{\Psi(\boldsymbol{\phi^*})} \approx \ket{GS}$. It is worth emphasizing that the optimization of the cost function $C(\boldsymbol{\phi})$ can only generate the ground state if the set of states spanned by $\ket{\Psi(\boldsymbol{\phi})}$ includes the target state $\ket{GS}$. For a very shallow circuit, this may not be the case. By increasing the depth of the circuit, the span set $\ket{\Psi(\boldsymbol{\phi})}$ will eventually be large enough to contain the target state $\ket{GS}$. The impact of the circuit depth on the span set $\ket{\Psi(\boldsymbol{\phi})}$ is schematically described in Fig.~\ref{fig:vqe_schematic}(b).

The required quantum resources can be quantified by the depth of the circuit. On the other hand, the most complex ingredient of a quantum circuit is the two-qubit gates, such as controlled-not. Therefore, one can simply use the number of two-qubit gates $R_Q$ as a quantification for quantum resources. For classical resources, we have to focus on the complexity of the optimization which depends on both the number of parameters $L$ and the number of iterations needed to converge to the minimum of the cost function. Therefore, a natural definition for classical resources can be  
\begin{equation}
	CR = \# \textmd{Iteration}  \times L 
\end{equation}
where $\# \textmd{Iteration}$ represents the number of iterations that a classical optimizer needs to converge. The classical resource $CR$ also quantifies the total number of measurements that one has to perform to fulfill the whole process. More precisely, the total number of measurements that one has to perform on the quantum simulator for the whole VQE simulation is twice $CR$. The factor $2$ is because for each parameter the gradient descent demands two successive measurements. 
Unlike $R_Q$ which only depends on the number of qubits and the number of circuit layers, $CR$ highly depends on the choice of the classical optimizer and initialization of the parameters $\boldsymbol{\phi}$. This is due to the fact that the number of iterations for the convergence of the optimizer depends on these two factors. In this paper, we use the gradient-based L-BFGS algorithm as our classical optimizer~\cite{Liu1989} and in order to have a sensible estimation for the number of iterations we repeat the procedure for $50$ different random initialization and take the average required iterations for computing $CR$. Note that the results do not depend on the choice of optimizer, however the convergence speed of the L-BFGS algorithm is normally faster than other gradient-based optimizers, such as Adam~\cite{kingma2014adam}. This can be attributed to the ability of L-BFGS to approximate the inverse Hessian matrix, incorporating second-order information that enables more informed and efficient updates during each iteration. In contrast, first-order methods, including Adam, depend solely on the gradient or first-order information for their updates. The faster convergence means that fewer iterations are needed for the optimization part, which results in the reduction of classical resources. 

Finally, it is worth emphasizing that the convergence of the VQE algorithm with average energy as a cost function can only be successful if energy gap $\Delta E$ between the ground and the first excited state is larger than the standard deviation of energy, namely $\Delta E > \sqrt{\langle \mathcal{H}^2 \rangle - {\langle \mathcal{H} \rangle}^2}$. If this condition is not satisfied, the classical optimizer cannot discriminate the two eigenstates and thus the VQE outcome will be an arbitrary superposition of them. One way to address this issue is to add extra terms to the cost function, based on the Hamiltonian symmetries whose values are different for the two eigenstates~\cite{lyu2022symmetry}. This will be further clarified in the following sections.

\begin{figure}[t]
	\centering
	\includegraphics[width=.9\columnwidth]{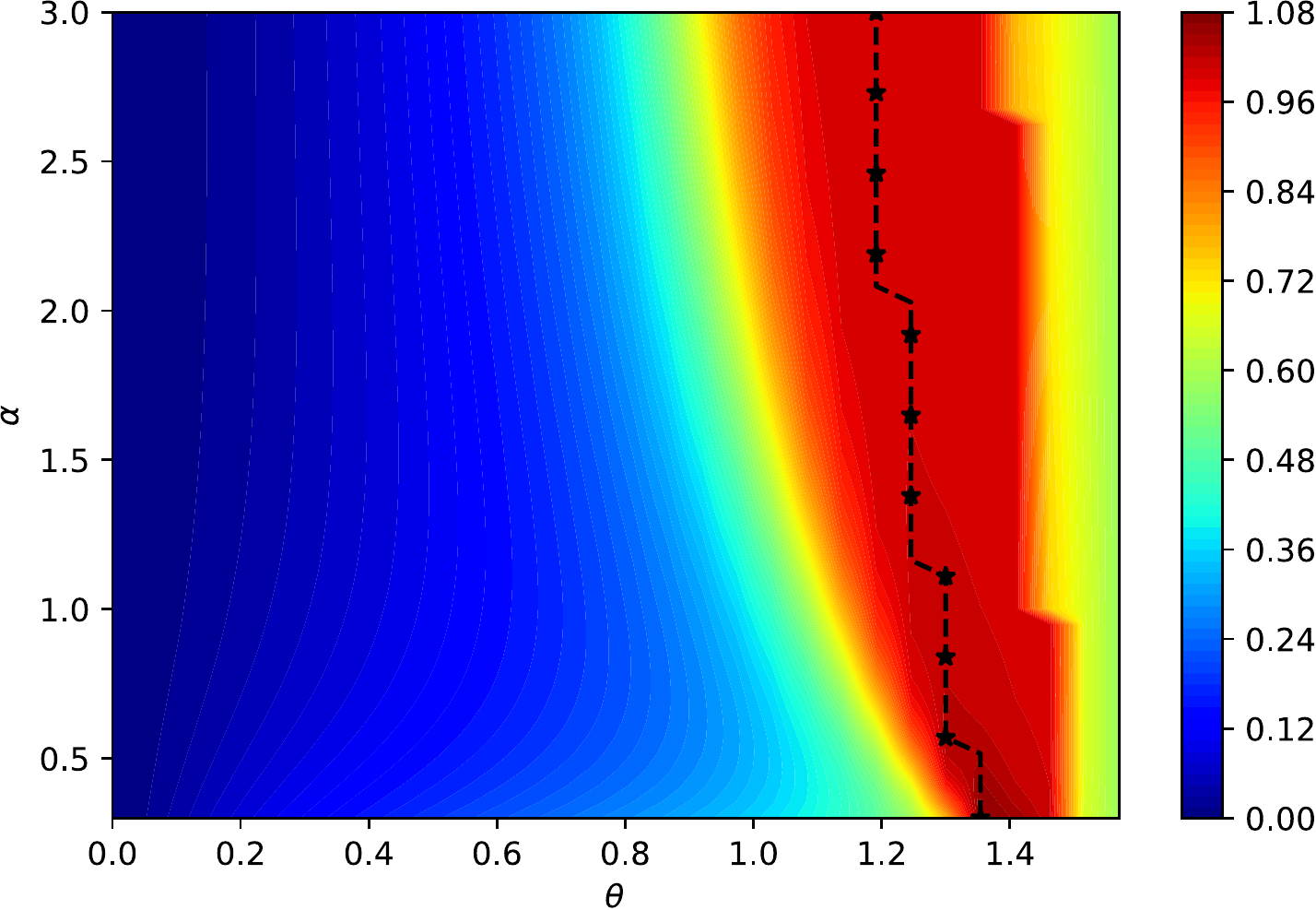}
	
	\caption{The  von Neumann entropy of the half system $S_V$, calculated from the ground state $\ket{GS}$ of long-range XY Hamiltonian with system size $N{=}12$, as a function of $\alpha$ and $\theta$. The peak of $S_V$, denoted by black dashed line, indicates the critical line as the phase boundary between the two phases. } 	
	\label{fig:phase_diagram}
\end{figure}

\section{Model}\label{section:model}

We study the VQE simulation for generating the ground state of long-range interacting system which is described by the Hamiltonian
\begin{equation}
	\mathcal{H}=\sum_{k=0}^{N-1} \mathcal{H}_k
\end{equation}
where $\mathcal{H}_0$ denotes the single particle interaction part of the Hamiltonian, e.g. interaction with external fields, and $\mathcal{H}_k$ (with $k>0$) represents the interaction between all qubits with distance $k$. In this paper, we consider long-range XY model in transverse field such that
\begin{eqnarray} 
	\mathcal{H}_0&=&\cos(\theta) \sum_{j=1}^N \sigma^z_j \cr
	\mathcal{H}_k&=&\sin(\theta) \sum_{j=1}^{N-k}  \frac{1}{2 k^\alpha} \left [ (1 + \gamma) \sigma^x_j \sigma^x_{j+k} + (1 - \gamma) \sigma^y_j \sigma^y_{j+k} \right ] \cr & & 
	\label{eq:Hamiltonian_Ising}
\end{eqnarray}
where $N$ represents the number of qubits, $\gamma$ is the anisotropy, $\mathcal{H}_k$ (with $k{\geq}1$) indicates the interaction between qubits and $0{\le} \theta {\le} \pi/2$ controls the phase of the system as $\cot(\theta)$ is the ratio between the exchange interaction and the transverse field. 
The strength of long-range interaction is controlled by exponent $\alpha$. The Hamiltonian in Eq.~\eqref{eq:Hamiltonian_Ising} covers a wide range of interactions as the parameters $\theta$, $\gamma$ and $\alpha$ vary. For instance, $\alpha=0$ represents a fully connected graph in which all the qubits interact equally and as $\alpha$ increases the interaction tends to become short ranged such that in the limit of $\alpha {\to} \infty$ the  Hamiltonian~\eqref{eq:Hamiltonian_Ising} becomes the conventional nearest neighbor XY model which can be solved analytically through Jordan-Wigner transformation~\cite{sachdev2011quantum}. Moreover, by controlling the anisotropy $\gamma$, one can change the Hamiltonian from Ising in transverse field, i.e. $\gamma=1$, to anisotropic XX model, i.e. $\gamma = 0$.

\begin{figure*}[t]
	\centering
	\includegraphics[width=2\columnwidth+\columnsep]{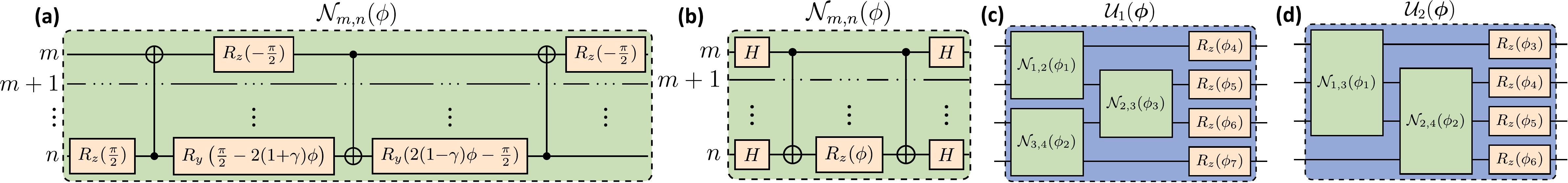}
	
	\caption{(a) A schematic picture of the circuit implementation of $\mathcal{N}_{i, j}(\phi)$. (b) A simplified circuit implementation (with only two controlled-not gate) of $\mathcal{N}_{i, j}(\phi)$ for the specific value of $\gamma{=}1$. (c) Circuit for realizing $\mathcal{U}_1(\boldsymbol{\phi})$ for a system of length $N{=}4$. (d) Circuit for realizing $\mathcal{U}_2(\boldsymbol{\phi})$ for a system of length $N{=}4$}	
	\label{fig:ansatz}
\end{figure*}

For any given $\alpha$ the long-range XY Hamiltonian~\eqref{eq:Hamiltonian_Ising} has a second order quantum phase transition in its ground state $\ket{GS}$ at a special value of $\theta=\theta_c$ which separates a paramagnetic phase from an antiferromagnetic phase. 
The paramagnetic phase ($\theta<\theta_c$) is characterized by all spins aligned in the transverse field direction. For short-range interacting systems, i.e. large $\alpha$, the antiferromagnetic phase can be characterized by staggered magnetization at a proper direction depending on $\gamma$~\cite{sachdev2011quantum}. In the case of long-range, characterization of the antiferromagnetic phase becomes more complex~\cite{koffel2012entanglement}. 
As for the case of nearest neighbor interaction, i.e.  $\alpha {\to} \infty$, the quantum phase transition is well-known to be at $\theta_c=\pi/4$. For $\theta\le \theta_c$ the system is in a paramagnetic phase with a unique ground state aligned by the external magnetic phase in $z$ direction. For $\theta > \theta_c$ the system is in an antiferromagnetic phase which becomes gapless in the thermodynamic limit.
The Hamiltonian~\eqref{eq:Hamiltonian_Ising} supports a special symmetry with $\mathcal{Z}=\otimes_{k=1}^N \sigma_k^z$ such that $[\mathcal{H},\mathcal{Z}]=0$. Consequently, each eigenstate of the Hamiltonian has an eigenvalue of $\pm 1$ with respect to $\mathcal{Z}$. In the antiferromagnetic phase, where the energy gap is small (in the thermodynamic limit it becomes zero), this symmetry can be used to enhance the quality of convergence for the VQE algorithm to distinguish between the two lowest energy eigenstates. In particular, in this paper, we focus on the ground state with $+1$ eigenvalue, namely $\mathcal{Z}\ket{GS}=\ket{GS}.$

In Ref.~\cite{koffel2012entanglement}, the phase diagram of the system for $\gamma=1$ (i.e. transverse Ising) as a function of $\theta$ and $\alpha$ has been explored through entanglement analysis using matrix product states for large system sizes $N$.  We perform similar analysis to short systems that we use for our VQE analysis to see how the behavior of the system approaches its thermodynamic limit. Without loss of generality, we also fix $\gamma=1$ and compute the ground state of the system for given values of $\alpha$ and $\theta$. One can trace out half of the qubits from the right side of the system to obtain the reduced density matrix of the left side as $\rho_L=Tr_R \ket{GS}\bra{GS} $, where $Tr_R$ denotes the partial trace. The entanglement between the two halves of the system is then quantified through the von Neumann entropy $S_V=-Tr\left[ \rho_L \log(\rho_L) \right]$. 
In Fig.~\ref{fig:phase_diagram} we plot $S_V$ as a function of $\theta$ and $\alpha$ for the long-range XY Hamiltonian with system size $N{=}12$. The von Neumann entropy peaks at the criticality and thus can be used as an indicator for  the quantum phase transition. Similar results can be found for other $\gamma$'s.

\section{Digital VQE simulation of the ground state} \label{section:simulation}

In this section, we focus on digital VQE simulation of the ground state of the Hamiltonian~\eqref{eq:Hamiltonian_Ising} across its phase diagram as $\theta$ and $\alpha$ vary. In order to prevent the complexity of degenerate ground states in the antiferromagnetic phase, we include the $\mathcal{Z}$ symmetry in the cost function of the system~\cite{lyu2022symmetry} such that 
\begin{equation}
	C(\boldsymbol{\phi}) = \langle \mathcal{H} \rangle + (\langle \mathcal{Z} \rangle - 1) ^ 2.
\end{equation}
The second term in the cost function penalizes the eigenstates whose $\mathcal{Z}$ eigenvalue is $-1$ and thus targets the desired ground state across the whole phase diagram.

\begin{figure*}[t]
	\centering
	\includegraphics[width=1.8\columnwidth]{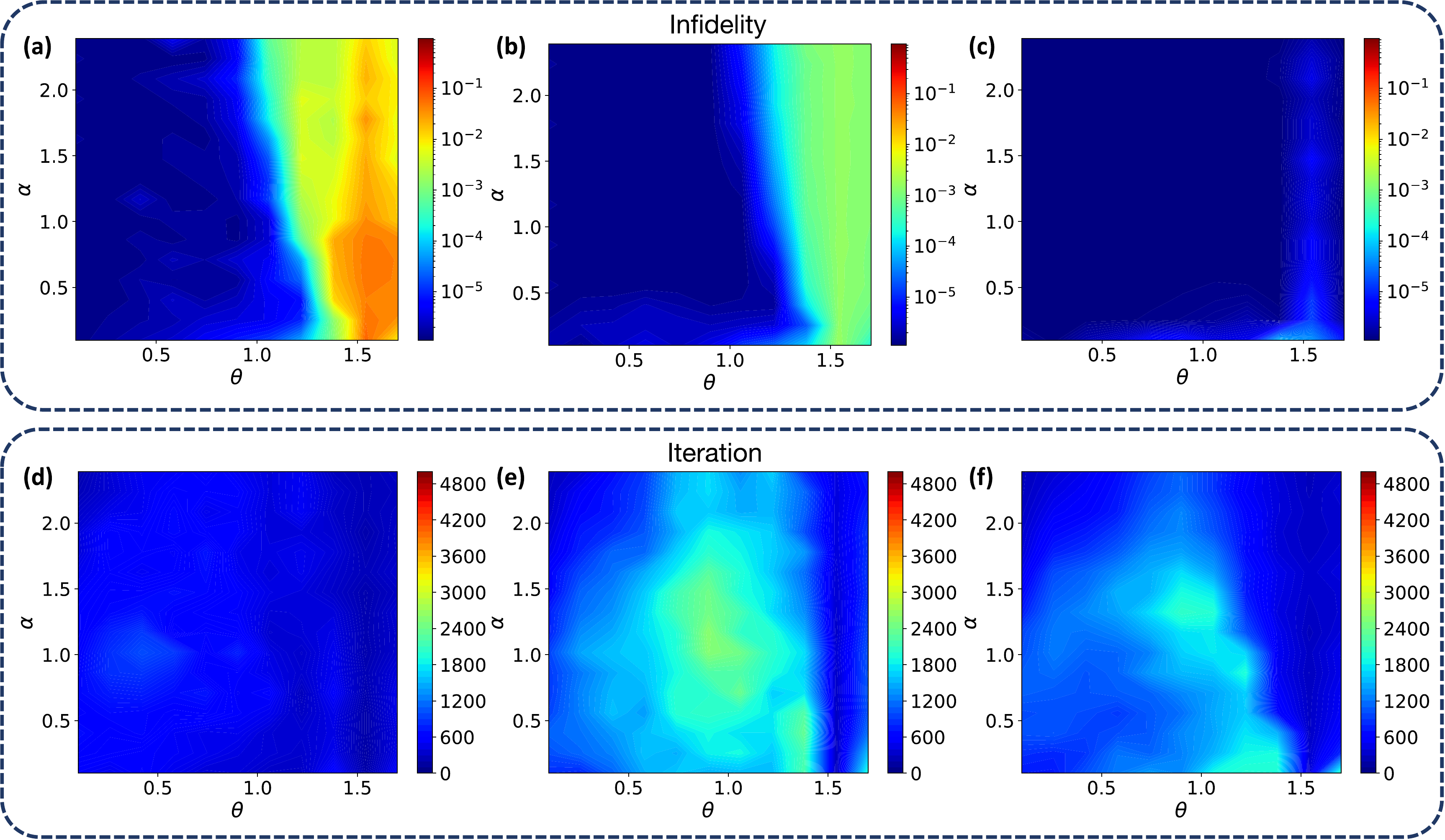}
	
	\caption{The VQE simulation for the ground state of the Hamiltonian $\mathcal{H}$ in a system of length $N=10$. The upper panel shows the infidelity $1{-}\mathcal{F}$ as a function of $\theta$ and $\alpha$ for the circuits of: (a) $2$ layers of the form $\mathcal{U}_1\mathcal{U}_1$; (b) $4$ layers of the form $\mathcal{U}_1\mathcal{U}_1\mathcal{U}_1\mathcal{U}_1$; and (c) $6$ layers of the form $\mathcal{U}_1\mathcal{U}_1\mathcal{U}_1\mathcal{U}_1\mathcal{U}_1\mathcal{U}_1$. The lower panel shows the corresponding  optimization iterations for: (d) $2$; (e) $4$; and (f) $6$ layers, and the number of parameters to be optimized are: $38$, $76$, $114$, respectively.}	\label{fig:C111}
\end{figure*}

Regarding the quantum simulator, we consider an array of qubits in a one-dimensional geometry with different levels of connectivity. The circuits in which one can perform two-qubit entangling gates, e.g. controlled-not gate, between a pair of qubits with distance $k$ (with $k\ge 1$) are called ansatz with $k$-distance connectivity. For instance, $1$-distance ($2$-distance) connectivity means that controlled-not gates are only allowed between nearest neighbor (and next nearest neighbor) qubits. 
As a two-qubit unitary operation between qubits $m$ and $n$, we use the following operation: 
\begin{equation}
	\mathcal{N}_{m,n}(\phi) =  e^{-i \phi \left [ (1 + \gamma)\sigma^x_m\sigma^x_{n} + (1 - \gamma)\sigma^y_m\sigma^y_{n} \right ] }
\end{equation}
which can be implemented by the circuit in Fig.~\ref{fig:ansatz}(a)~\cite{Vatan2004}. As the figure shows, to implement this operation one needs three controlled-not gates. However, one can simplify the circuit to only use two controlled-not gates for the specific value of $\gamma{=}1$, as shown in Fig.~\ref{fig:ansatz}(b). The two qubit unitary operations $\mathcal{N}_{m,n}(\phi)$ is used to generate a VQE ansatz inspired by quantum approximate optimization algorithm~\cite{farhi2014quantum}. Depending on the connectivity between the qubits, one can apply $\mathcal{N}_{m,n}(\phi)$ between qubits of different distances. 
The VQE circuit with $k$-distance can be described by the following unitary operation.
\begin{equation}
	\mathcal{U}_k (\boldsymbol{\phi}) =  R_z(\phi_1^{(0)},\cdots,\phi_N^{(0)}) 
	\text{   } \prod_{j} \mathcal{N}_{j,j+k}(\phi_j^{(k)})
\end{equation}  
where  $\boldsymbol{\phi}=(\phi_1^{(0)},\cdots,\phi_N^{(0)},\phi_1^{(k)}, \cdots,\phi_{N-k}^{(k)})$ are the tunable angles and $R_z(\phi_1^{(0)},\cdots,\phi_N^{(0)})$ is a set of rotation gates acting on each qubit as
\begin{equation}
	R_z(\boldsymbol{\phi^{(0)}})=\bigotimes_{j=1}^N \begin{bmatrix} 
		e^{-i \phi_j^{(0)}/2} & 0 \\
		0 & e^{i \phi_j^{(0)}/2} \\
	\end{bmatrix}.
\end{equation} 
Therefore, each $\mathcal{U}_k(\boldsymbol{\phi})$ contains $L_k=2N-k$ parameters to be optimized. This includes $N-k$ two-qubit parameters $\phi_j^{(k)}$ and $N$ one-qubit parameters $\phi_j^{(0)}$ for local rotations. 
In Figs.~\ref{fig:ansatz}(c)-(d) we schematically depict the circuit for $\mathcal{U}_1$ and $\mathcal{U}_2$, respectively. 
Note that, the blue boxes in Figs.~\ref{fig:ansatz}(c)-(d) represent one layer of the circuit. In order to converge to the ground state, we need to concatenate several of these layers. The number of layers $M$ is chosen such that the ground state is achieved with high accuracy. 
Quantum circuits for $\mathcal{U}_k$ (with $k>2$) can be similarly produced. In general, any single layer with a quantum circuit which implements $\mathcal{U}_k$ contains $3(N-k)$ controlled-not gates. For the specific value of $\gamma{=}1$, using the simplified circuit implementation, each layer will instead contain $2(N-k)$ controlled-not gates.
Depending on the connectivity of qubits different combinations of $\mathcal{U}_k$'s can be used as our VQE circuit. 
Note that, all the VQE simulations are performed using the simulator MindQuantum~\cite{mq_2021} with the choice of the classical optimizer being the L-BFGS algorithm~\cite{Liu1989}. 
Moreover, we apply the layer-recursive method~\cite{Lyu2020accelerated} for simulating circuits with the number of layers $M {>} 1$ which will start the training by only one layer then a new layer will be added after the existing circuit is optimized. 
The initial parameters for the first layer are randomly sampled from a normal distribution. When a new layer is added, the optimized parameters from the last layer of the preceding circuit serve as the initial parameters for the subsequent layer.
This method has been proven to provide a great improvement for the convergence of the VQE optimization~\cite{Lyu2020accelerated}.

We first focus on $1$-distance connectivity. A circuit of $M$ layers is described by $U_M=\prod_{l=1}^{M} \mathcal{U}_1(\boldsymbol{\phi_l})$, where  $\boldsymbol{\phi_l}$ represents the tunable parameters at layer $l$. One can quantify the performance of the VQE through fidelity of the real ground state and the output of the trained quantum simulator $\mathcal{F}{=}|\langle GS| \Psi(\boldsymbol{\phi}) \rangle|^2$. 
In Fig.~\ref{fig:C111}(a)-(c),  we plot the infidelity $1-\mathcal{F}$ for simulating the ground state of the Hamiltonian~\eqref{eq:Hamiltonian_Ising} with $\gamma=1$ (i.e. transverse Ising model) in a system of length $N=10$ across its phase diagram for $M=2$, $M=4$ and $M=6$ layers, respectively. 
As the number of circuit layer increases, it is clear from the figures that the performance of the VQE improves significantly. In particular, it converges to a low infidelity in most of the phase diagram with the circuit depth of $M=6$. However, it fails to yield the ground state with low infidelity when $\alpha$ is less than $0.5$ where the Hamiltonian tends to be fully connected. The optimizer computes the iterations until the improvement of infidelity is less than a threshold, here $\sim 10^{-9}$. We choose this small threshold to make sure that the optimizer truly saturates. In Figs.~\ref{fig:C111}(d)-(f), we plot the corresponding optimization iterations needed for convergence to the obtained infidelities. 
An interesting feature can be seen in Fig.~\ref{fig:C111}(f), whose corresponding infidelity in Fig.~\ref{fig:C111}(c) shows good  performance in the entire phase diagram. In fact, the required iteration is larger along the phase transition line, in particular when $\alpha$ is small (i.e. the system is more long-ranged). This can be attributed to the complexity of the ground state at criticality which naturally exhibits more entanglement and long-range correlations.

\begin{figure}[t]
	\centering
	\includegraphics[width=\columnwidth]{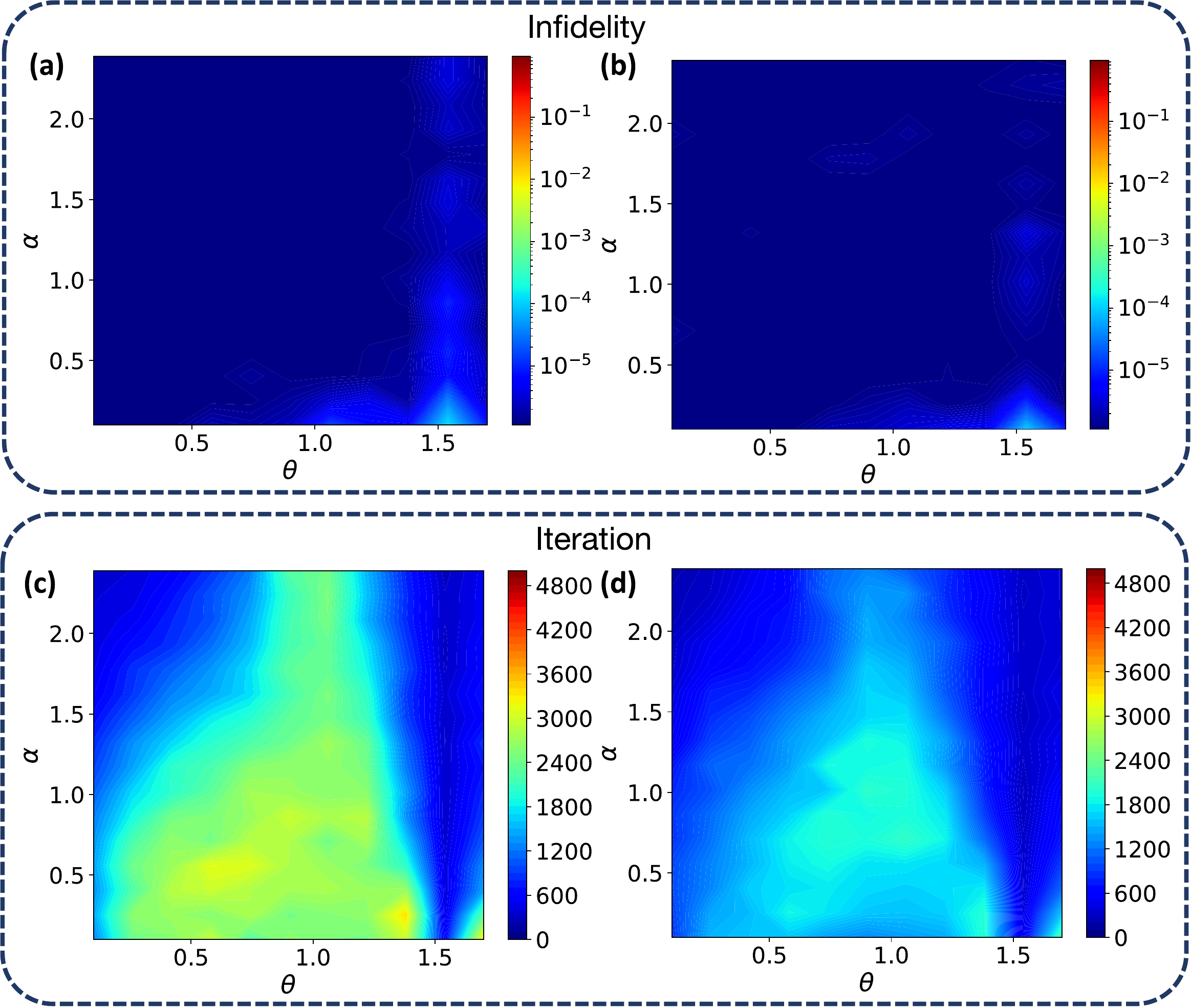}
	
	\caption{The VQE simulation for the ground state of the Hamiltonian $\mathcal{H}$ in a system of length $N=10$. The upper panel shows the infidelity $1-\mathcal{F}$ as a function of $\theta$ and $\alpha$ for the circuits of: (a) $6$ layers of the form $\mathcal{U}_1\mathcal{U}_1\mathcal{U}_1\mathcal{U}_2\mathcal{U}_2\mathcal{U}_2$; and (b) $6$ layers of the form $\mathcal{U}_1\mathcal{U}_1\mathcal{U}_2\mathcal{U}_2\mathcal{U}_3\mathcal{U}_3$. The lower panel shows the corresponding  optimization iterations for: (c) $\mathcal{U}_1\mathcal{U}_1\mathcal{U}_1\mathcal{U}_2\mathcal{U}_2\mathcal{U}_2$; and (d) $\mathcal{U}_1\mathcal{U}_1\mathcal{U}_2\mathcal{U}_2\mathcal{U}_3\mathcal{U}_3$ layers, and the number of parameters to be optimized are: $111$, $108$, respectively. }	
	\label{fig:C2121}
\end{figure}

One would naturally wonder if higher distance connectivity between the qubits could be of any benefits for VQE simulation of the Hamiltonian~\eqref{eq:Hamiltonian_Ising}. 
To explore this, we consider a circuit with both $2$- and $1$- distance connectivity. Unlike the case of $1$-distance connectivity, in this case the order of the layers is another degree of freedom which can be optimized to get better performance. To investigate different configurations, we keep $M=6$ layers for the transverse Ising model (i.e. $\gamma{=}1$) of size $N=10$ qubits. Later, we will consider other $\gamma$'s too. We first consider alternating layers, implemented by $\mathcal{U}_1$ and $\mathcal{U}_2$. Our simulation shows that the performance is obtained for the configuration $\mathcal{U}_1\mathcal{U}_1\mathcal{U}_1\mathcal{U}_2\mathcal{U}_2\mathcal{U}_2$. For the sake of brevity, we only show the results for this optimal configuration in Fig.~\ref{fig:C2121}(a) and Fig.~\ref{fig:C2121}(c) for the achieved infidelity $1{-}\mathcal{F}$ and the corresponding iterations, respectively. The infidelity shows improvement in comparison with the circuit using only $\mathcal{U}_1$ with the same number of layers, see Fig.~\ref{fig:C111}(c). For the case of $3$-distance connectivity, our simulations show that the best performance can be obtained using $\mathcal{U}_1\mathcal{U}_1\mathcal{U}_2\mathcal{U}_2\mathcal{U}_3\mathcal{U}_3$. The infidelity and its corresponding iteration is shown in Fig.~\ref{fig:C2121}(b) and Fig.~\ref{fig:C2121}(d). The overall improvement of infidelity in compare to the circuit with only $1$-distance gates, see Fig.~\ref{fig:C111}(c), and the circuit with $1$- and $2$-distance gates, see Fig.~\ref{fig:C2121}(a), is evident. 
In order to have a quantitative comparison between these circuit designs with different qubit distance connectivity, in Table.~\ref{table:gamma1_Fmin} we compare the minimum achievable fidelity $\mathcal{F}_{min}$, the average achievable fidelity $\mathcal{F}_{avg}$, and the maximum classical resources $CR_{max}$ across the entire phase diagram. Additionally, the number of two-qubit gates $R_Q$ is included as a metric for assessing quantum resources. Given that $\mathcal{F}{\approx}1$ can be readily achieved with smaller $\theta$ values in the phase diagram, the maximum achievable fidelity $\mathcal{F}_{max}$ is omitted in this discussion. Indeed, the information shown in Table.~\ref{table:gamma1_Fmin} provides a comparison over the worst and the average performance of different circuit designs in the entire phase diagram.
Clearly, the circuit represented by $\mathcal{U}_1\mathcal{U}_1\mathcal{U}_2\mathcal{U}_2\mathcal{U}_3\mathcal{U}_3$ achieves higher fidelity even at its worst point of the phase diagram. Interestingly, the number of controlled-not gates $R_Q$ is also minimum for this circuit configuration, showing the minimal demand of quantum resources. Similarly, the maximum classical resources $CR_{max}$ in the entire phase diagram is far less for this circuit compared to the others.

\begin{table}[h!] 
	\centering
	
\begin{tabular}[c]{ |K{1.2cm}|K{1.2cm}|K{1.2cm}|K{1.2cm}|K{1.2cm}|}
	\hline
	
	\multicolumn{5}{|c|}{6 layers } \\
     \hline
	Circuit & $\mathcal{F}_{min}$ & $\mathcal{F}_{avg}$ & $CR_{max}$ & $R_Q$ \\
	\hline 
	111111 & 0.594 & 0.974 & 250528 & 108 \\[1pt]
	212121 & 0.651 & 0.981 & 412569 & 102 \\[1pt]
	222111 & 0.680 & 0.985 & 372136 & 102 \\[1pt]
	321321 & 0.624 & 0.979 & 355488 & 96 \\[1pt]
	332211 & 0.803 & 0.992 & 238710 & 96 \\
	\hline
\end{tabular}
	\caption{
	The minimum fidelity $\mathcal{F}_{min}$, and the average fidelity $\mathcal{F}_{avg}$ achievable across the entire phase diagram for various circuit designs with $6$ layers are presented, along with the maximum required classical resources, $CR_{max}$. Additionally, the number of two-qubit gates $R_Q$ is provided for each specific circuit design. Each circuits $i_1 i_{2} \cdots i_k$ represents a quantum circuit with unitary operation of the type  $\mathcal{U}_{i_k} \cdots \mathcal{U}_{i_2} \mathcal{U}_{i_1}$.}
	\label{table:gamma1_Fmin}
\end{table}

If we restrict ourselves to $1$- and $2$-distance gates, we also see that the overall performance of $\mathcal{U}_1\mathcal{U}_1\mathcal{U}_1\mathcal{U}_2\mathcal{U}_2\mathcal{U}_2$ outperforms the other configurations such as $\mathcal{U}_1\mathcal{U}_2\mathcal{U}_1\mathcal{U}_2\mathcal{U}_1\mathcal{U}_2$. This can be seen in Table.~\ref{table:gamma1_Fmin}. While the minimum fidelities $\mathcal{F}_{min}$ are pretty close to each other, the classical resources $CR_{max}$ for the former circuit is much less. This shows that by grouping the layers with long-distance connectivity at the beginning of the circuit, one can achieve better overall performance. 
Indeed, such circuit design is capable for creating long-distance correlations between distant qubits using fewer layers. This is particularly useful for long-range interacting systems which inherently have larger values of long-range correlations. As a result, this specific design of the circuit achieves better performance for simulating the ground state of long-range interacting systems.

\section{Scaling up} \label{section:scaling}

\begin{figure}[t]
	\centering
	\includegraphics[width=0.8\columnwidth]{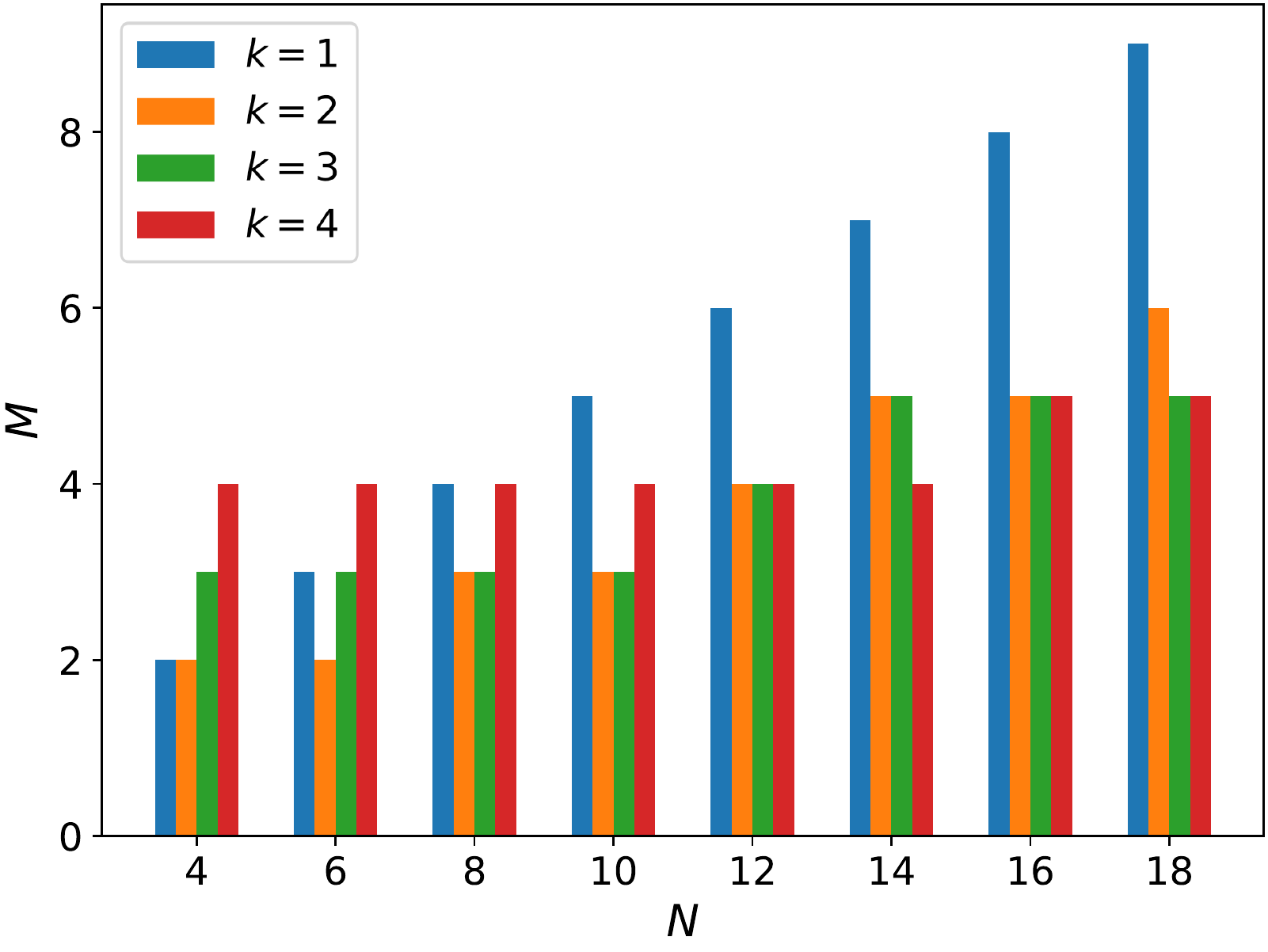}
	
	\caption{The required circuit layer $M$ to obtain fidelity $\mathcal{F}{\geq}0.95$ for simulating the ground state of the transverse Ising Hamiltonian (i.e. $\gamma{=}1$) with $\alpha{=}0.5$ and $\theta{=}0.425\pi$, in various system sizes $N$ as well as different maximum qubit connectivities $k$.}	
	\label{fig:bar_fig}
\end{figure}

In this section, we focus on the impact of qubit connectivity for simulation of larger system sizes. In particular, we consider the optimal circuit design in which the layers with the same qubit connectivity are grouped together and positioned in descending order. For instance, in a circuit with $M=5$ layers and the maximum qubit connectivity $k{=}3$, we use $\mathcal{U}_1 \mathcal{U}_2 \mathcal{U}_2 \mathcal{U}_3 \mathcal{U}_3$. For a transverse Ising chain (i.e. $\gamma{=}1$) with $\alpha{=}0.5$, which represents a strong long-range interaction, and $\theta{=}0.425\pi$, which is close to the critical point, we simulate the ground state of the Hamiltonian to obtain fidelity $\mathcal{F}\geq 0.95$. This choice of $\alpha$ and $\theta$ represents the most difficult part of the phase diagram for the VQE convergence. By fixing qubit connectivity $k$, we determine how many circuit layers $M$ is required to achieve the fidelity $\mathcal{F}\geq 0.95$. The results for various system sizes are shown in Fig.~\ref{fig:bar_fig}. As one can see, for a given qubit connectivity $k$, at least $k$ layers are required. This guarantees that layers with all available connectivities, namely $k$, $k-1$, $\cdots$, $1$, are required for achieving high fidelity VQE performance. While large $k$ does not show any advantage for small system sizes, its benefit becomes very evident for large systems. In fact, when $k$ is larger, the VQE procedure converges with fewer layers as the system size increases. This shows that having larger qubit connectivity is hugely beneficial for scaling up the VQE simulation of long-range interacting systems.

\section{Generality with respect to \texorpdfstring{$\gamma$}{gamma}} \label{section:generality}

So far, we have focused on transverse Ising model with $\gamma{=}1$. In this section, we show that our strategy for circuit design is applicable for other values of $\gamma$. We have to note that for $\gamma{\neq}1$ the circuit design for the two-qubit gate operation $\mathcal{N}_{m,n}(\phi)$ has to change as shown in Fig.~\ref{fig:ansatz}(a). 
For a chain of size $N{=}10$ and a circuit with $M{=}6$ layers, we present the minimum achievable fidelity $\mathcal{F}_{min}$, average achievable fidelity $\mathcal{F}_{avg}$, and maximum classical resources $CR_{max}$, obtained across the entire phase diagram, in Table.~\ref{table:Fmin} for three different values of $\gamma$. As $\mathcal{F}{\approx}1$ can be readily attained with smaller $\theta$ values in the phase diagram, we have excluded the maximum achievable fidelity $\mathcal{F}_{max}$ from this discussion.
The superiority of the circuit $\mathcal{U}_1\mathcal{U}_1\mathcal{U}_2\mathcal{U}_2\mathcal{U}_3\mathcal{U}_3$ is evident in achieving the best fidelity $\mathcal{F}_{min}$, as the worst outcome of the circuit in the entire phase diagram. Remarkably, this circuit also demands less classical resources in comparison with other circuit designs. This shows that grouping the layers with long-distance connectivity at the beginning of the circuit can significantly enhance the performance of the VQE in terms of both fidelity and required classical resources.

\begin{table*}[t] 
	\centering
\begin{tabular}[c]{ |K{1.2cm}|K{1.2cm}|K{1.2cm}|K{1.2cm}|K{1.2cm} |K{1.2cm}|K{1.2cm}|K{1.2cm}|K{1.2cm}|K{1.2cm}|}
	\hline
	
	& \multicolumn{3}{c|}{$\gamma=0$} & \multicolumn{3}{c|}{$\gamma=0.5$} & \multicolumn{3}{c|}{$\gamma=1$} \\
     \hline
	Circuit& $\mathcal{F}_{min}$ & $\mathcal{F}_{avg}$  & $CR_{max}$ & $\mathcal{F}_{min}$ & $\mathcal{F}_{avg}$  & $CR_{max}$ & $\mathcal{F}_{min}$ & $\mathcal{F}_{avg}$  & $CR_{max}$ \\
	\hline
	111111 & 0.38 & 0.959 & 73019 & 0.51 & 0.985 & 185799 & 0.594 & 0.974 & 250528 \\[1pt]
	212121 & 0.81 & 0.997 & 461666 & 0.74 & 0.991 & 546297 & 0.651 & 0.981 & 412569\\[1pt]
	222111 & 0.54 & 0.964 & 400336 & 0.46 & 0.979 & 582234 & 0.680 & 0.985 & 372136 \\[1pt]
	321321 & 0.53 & 0.964 & 382494 & 0.49 & 0.983 & 507185 & 0.624 & 0.979 & 355488 \\[1pt]
	332211 & 0.79 & 0.995 & 341843 & 0.86 & 0.994 & 397260 & 0.803 & 0.992 & 238710 \\
	\hline
\end{tabular}
	\caption{
	The obtainable minimum fidelity $\mathcal{F}_{min}$, average fidelity $\mathcal{F}_{avg}$, and maximum necessary classical resources $CR_{max}$ across the whole phase diagram with different designs of the circuit with $6$ layers, for three different values of $\gamma$. Each circuits $i_1 i_{2} \cdots i_k$ represents a quantum circuit with unitary operation of the type  $\mathcal{U}_{i_k} \cdots \mathcal{U}_{i_2} \mathcal{U}_{i_1}$.}
	\label{table:Fmin}
\end{table*}

\begin{figure}[t]
	\centering
	\includegraphics[width=\columnwidth]{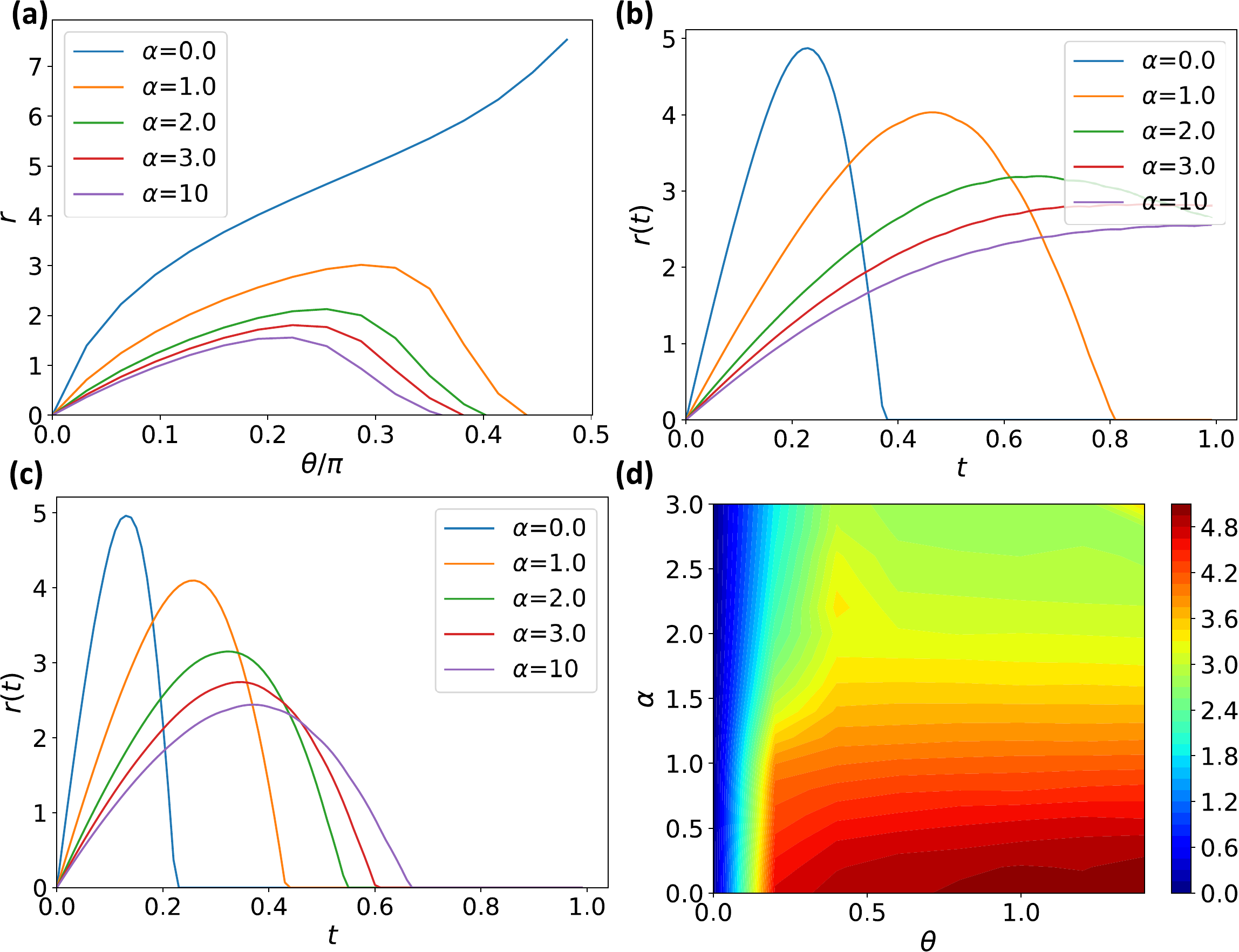}
	
	\caption{ (a) the ground state spin squeezing $r$ as a function of $\theta$ for different values of $\alpha$. (b) The spin squeezing $r$ as a function of time for $\theta=\pi/8$ for various choices of $\alpha$. (c) The spin squeezing $r$ as a function of time for $\theta=\pi/4$ for various choices of $\alpha$. (d) The maximum obtainable squeezing $r(t^*)$ as a function of $\alpha$ and $\theta$. } 	\label{fig:squeezing}
\end{figure}

\section{Spin squeezing} \label{section:squeezing}

Generating the eigenstates of a Hamiltonian is not the only application of variational quantum algorithms~\cite{cerezo2021variational}. In principle, any problem that can be written variationally in terms of minimization of a cost function can be solved on a quantum simulator. This has led to a wide range of variational quantum algorithms for solving problems in linear algebra~\cite{xu2021variational,endo2020variational}, combinatorial optimization~\cite{farhi2014quantum,liu2022layer} and machine learning~\cite{biamonte2017quantum,arunachalam2017survey,ciliberto2018quantum,dunjko2018machine,farhi2018classification,schuld2019quantum}.  
In this section, we show that our circuits with $k$-distance qubit connectivity can indeed provide advantage for producing an important class of states with a non-classical feature called spin squeezing.

Spin squeezed quantum states~\cite{kitagawa1993squeezed,ma2011quantum} are an important class of entangled states which have fundamental and practical applications. Fundamentally, these states can be used for testing the foundations of quantum mechanics through violating Bell inequalities~\cite{schmied2016bell,engelsen2017bell}. From a practical perspective, these states can be regarded as resource for quantum metrology protocols which can achieve quantum enhanced sensitivity~\cite{pezze2018quantum}. Spin squeezing is defined for a set of spin particles whose collective spin has a net polarization in one direction while the fluctuations of the corresponding transverse spin component 
is suppressed. We define the collective spin components in $x$, $y$ and $z$ directions as $S_{\text{tot}}^{x,y,z}=1/2\sum_k \sigma^{x,y,z}_k$. Assuming that the net spin polarization is in the $z$ direction one can quantify spin squeezing through
\begin{equation}\label{eq:spin_squeezing_log}
	r=\text{max} \left[ -10 \text{ } \log_{10} \xi^2 ,0  \right]
\end{equation}
in which the parameter $\xi^2$ is defined as~\cite{wineland1994squeezed}
\begin{equation}\label{eq:spin_squeezing}
	\xi^2=\frac{N \text{ }\text{min}_{\beta} \text{Var}\left[ \cos(\beta) S_{\text{tot}}^x + \sin(\beta)S_{\text{tot}}^y \right]}{|\langle S_{\text{tot}}^z \rangle|^2}
\end{equation}  
where, $\text{Var}[\bullet]$ stands for variance and angle $\beta$ spans the collective spin in the transverse direction. To have nonzero squeezing $r$, one has to reach $\xi^2<1$, which requires strong spin polarization $|\langle S_{\text{tot}}^z \rangle|^2$ along the $z$ direction and suppressed variance for the collective spin component in the transverse direction. 

In practice, creating spin-squeezed states is very challenging. Here, we consider three different methods for generating such states. The first approach is to search for spin squeezing in the ground state of long-range Ising Hamiltonian in the transverse field. The second approach is to use the quench dynamics of such Hamiltonian for creating spin squeezed states. Finally, the third approach is to use a variational quantum circuit to generate such states using the parameter $\xi^2$, in Eq.~\eqref{eq:spin_squeezing}, as the cost function~\cite{kaubruegger2019variational,marciniak2021optimal}. 

Let's consider $N$ spin-1/2 particles interacting via transverse Ising Hamiltonian~\eqref{eq:Hamiltonian_Ising} with $\gamma{=}1$. For any given values of $\alpha$ and $\theta$ one can compute the squeezing parameter $r$ for the ground state $\ket{GS}$. In Fig.~\ref{fig:squeezing}(a), we plot the spin squeezing $r$ as a function of $\theta$ in the whole phase diagram, for various values of $\alpha$ in a system of $N=10$ particles. Two main features can be observed: (i) decreasing $\alpha$ (i.e. making the interaction more long-ranged) monotonically enhances the spin squeezing; (ii)  for a fixed $\alpha$ the spin squeezing takes its maximum around the phase transition point. It is worth emphasizing that the minimum of $\xi^2$, in Eq.~\eqref{eq:spin_squeezing}, takes place at $\beta=0$ in the whole phase diagram for the ground state squeezing.

Although, the ground state of the system reveals squeezing, its value can be enhanced through quench dynamics~\cite{comparin2022robust}. In this case, we initialize the system in the state $\ket{\Psi(0)}=\ket{0,0,\cdots,0}$. In general, this state is not an eigenvector of the transverse Ising Hamiltonian~\eqref{eq:Hamiltonian_Ising} and thus evolves to $\ket{\Psi(t)}=e^{-iHt}\ket{\Psi(0)}$. By computing the squeezing parameter $r$ with respect to quantum state $\ket{\Psi(t)}$ one can explore the dynamics of spin squeezing as a function of time. In Figs.~\ref{fig:squeezing}(b)-(c) we plot the squeezing parameter $r(t)$ as a function of time for various choices of $\alpha$ when $\theta$ is fixed to $\theta=\pi/8$ and $\theta=\pi/4$, respectively. As the figures show $r$ starts from $0$ and  increases by time to reach its maximum at an optimal time $t=t^*$ and then decreases again. The maximum spin squeezing $r(t^*)$ quantifies the capacity of the Hamiltonian for dynamical production of spin squeezing.  In Fig.~\ref{fig:squeezing}(d), we plot the maximum obtainable squeezing $r(t^*)$ as a function of $\theta$ and $\alpha$ for a system of length $N=10$. Unlike the ground state, the optimal $\beta$ which minimizes $\xi^2$, in Eq.~\eqref{eq:spin_squeezing}, varies in the dynamical case as $\theta$ and $\alpha$ change.  It should be noted that using the dynamical approach results in more squeezing than the ground state of the same Hamiltonian, literally throughout the whole phase diagram.

As an alternative to the ground state or the dynamics of the transverse Ising Hamiltonian~\eqref{eq:Hamiltonian_Ising} with $\gamma{=}1$ for generating spin squeezed states, one can directly apply a variational quantum algorithm with $\xi^2$, given in Eq.~\eqref{eq:spin_squeezing}, as the cost function. Without loss of generality we fix $\beta=0$. We use the same quantum circuits which we have used for simulating the ground state.  The obtainable squeezing $r$ from a circuit of $M{=}6$ layers with different designs and the initial state $\ket{0,0,\cdots,0}$ is shown in Table.~\ref{table:squeezing}. 
For each circuit, the obtainable squeezing $r$ is averaged over $100$ repetition of the protocol and the uncertainty in the values of Table.~\ref{table:squeezing} is determined through the standard deviation. 
Interestingly, the circuits which perform better for simulating the ground state also provide higher squeezing rate $r$. In addition, variational quantum approach for generating spin squeezed states results in higher squeezing than both the ground state and the dynamics of long-range interacting Hamiltonian across the whole phase diagram. For instance, the highest squeezing rate $r$ which one can obtain by the ground state and the dynamics in the whole phase diagram is $r=4.4$ and $r=5.1$, respectively. Indeed, as Table.~\ref{table:squeezing} shows, one can outperform both of these strategies using variational quantum algorithms.

\begin{table}[h!] 
	\centering
	\begin{tabular}{ |c|c|c|c| } 
		\hline
		Circuit & 111111 & 222111 & 332211\\
		\hline 
		$r$ & $6.50 \pm 1.64$ & $6.83 \pm 1.39$ & $7.20\pm 0.94$ \\ 
		\hline
	\end{tabular} 
\caption{The squeezing which can be obtained from different VQA circuits with $6$ layers. Each circuit $i_1 i_2 i_3 i_4 i_5 i_6$ represents a quantum circuit of the type $\mathcal{U}_{i_6}\mathcal{U}_{i_5}\mathcal{U}_{i_4}\mathcal{U}_{i_3}\mathcal{U}_{i_2}\mathcal{U}_{i_1}$. }
\label{table:squeezing}
\end{table}

\section{Conclusion} \label{section:conclusion}

The most promising approach to achieve quantum advantage with imperfect NISQ simulators is variational quantum algorithm. VQE, as one the most widely used VQAs, has been developed for generating the ground state of many-body systems on current quantum simulators. In this paper, we have focused on digital VQE simulation of a large class of long-range Hamiltonians with a quantum phase transition. 
We find out that by making the interaction more long-ranged, the VQE algorithm results in lower fidelities and demands more optimization iterations.
The situation becomes worse as the Hamiltonian gets closer to its criticality. Interestingly, our simulations show that qubit connectivity in  quantum simulators has a direct impact on the performance of VQE. With the possibility of applying two-qubit entangling gates between distant qubits the achievable fidelity is improved, in particular, for Hamiltonians which are more long-ranged. 
Interestingly, such enhancement in fidelity is combined with improvement in demanding both quantum and classical resources. 
The performance is further enhanced if the layers with larger distance connectivity grouped together and act before the layers with shorter distance connectivity.

As an alternative application, we have shown that the same design of circuits for variationally generating spin squeezed states, as resource for quantum metrology. The variational generation of spin squeezing outperforms the results from the ground state and quench dynamics in long-range interacting systems. 

Current NISQ simulators provide various forms of qubit connectivity. In certain physical setups, such as superconducting devices, the qubit connectivity is determined by fabrication and is usually limited to nearest neighbors. In other systems, such as ion-traps and Rydberg atoms, the connectivity is more versatile and operations between distant qubits are also possible. This suggests that ion-traps and Rydberg atoms are more suitable for simulating long-range interacting systems.

\section{Data availability}
All the codes for training the circuits are available online~\cite{Chufan2022long}, and the data presented on this paper will be provided upon reasonable requests from the authors.

\section{Acknowledgments}
A.B. acknowledges support from the National Key
R\&D Program of China (Grant No.2018YFA0306703),
the National Science Foundation of China (Grants
No. 12050410253, No. 92065115 and No. 12274059), and the Ministry of Science and Technology of China (Grant No.
QNJ2021167001L).

\bibliographystyle{apsrev4-1}
\bibliography{VQE_Long_Range}

\end{document}